\documentclass[12pt]{article}
\pdfoutput=1
\usepackage{amssymb,cite,graphicx}
\usepackage{slashed}
\usepackage{amsmath,bm,bbm}
\usepackage{amsfonts}
\usepackage[titletoc,title]{appendix}
\usepackage[small]{caption}
\usepackage[margin=1in]{geometry}
\usepackage[multiple]{footmisc}
\usepackage{mathtools}
\usepackage{slashed}
\usepackage[nottoc]{tocbibind}
\usepackage{xcolor}

\definecolor{mediumblue}{rgb}{0,0,0.8}
\usepackage{hyperref}
\hypersetup{
  linktocpage=true,
  colorlinks=true,
  citecolor=mediumblue,
  filecolor=mediumblue,
  linkcolor=mediumblue,
  urlcolor=mediumblue,
}

\graphicspath{{./}{./figures/}}
\numberwithin{equation}{section}
\allowdisplaybreaks%

\newcommand{\be}{\begin{equation}}
\newcommand{\ee}{\end{equation}}
\newcommand{\bea}{\begin{eqnarray}}
\newcommand{\eea}{\end{eqnarray}}

\begin{document}

\begin{titlepage}
  \begin{flushright}
 CERN-TH-2020-105
  \end{flushright}
  \medskip

  \begin{center}
    {\Large\bf\boldmath
    Exothermic Dark Matter for XENON1T Excess \vspace{0.3cm} \\
    }
\vspace{1.5cm}

    {\bf Hyun Min Lee }

    {\it Department of Physics, Chung-Ang University, Seoul
      06974, Korea}\\[0.2cm]
          {\it  CERN, Theory department, 1211 Geneva 23, Switzerland}\\[0.2cm]
  \end{center}

  \bigskip

  \begin{abstract}
    \noindent
    Motivated by the recent excess in the electron recoil from XENON1T experiment, we consider the possibility of exothermic dark matter, which is composed of two states with mass splitting. 
   The heavier state down-scatters off the electron into the lighter state, making an appropriate recoil energy required for the Xenon excess even for the standard Maxwellian  velocity distribution of dark matter. Accordingly, we determine the mass difference between two component states of dark matter to the peak electron recoil energy at about $2.5\,{\rm keV}$ up to the detector resolution, accounting for the recoil events over $E_R=2-3\,{\rm keV}$, which are most significant. 
   We include the effects of the phase-space enhancement and the atomic excitation factor to calculate the required scattering cross section for the Xenon excess. We discuss the implications of dark matter interactions in the effective theory for exothermic dark matter and a massive $Z'$ mediator and provide microscopic models realizing the required dark matter and electron couplings to $Z'$.

  \end{abstract}

\vspace{4.5cm}
  \begin{flushleft}
    Email: hminlee@cau.ac.kr 
  \end{flushleft}
\end{titlepage}
 
\section{Introduction}

The nature of dark matter has been a long standing mystery for astrophysics and particle physics. 
Weakly Interacting Massive Particles (WIMPs) have been searched for in various direct detection, cosmic-ray as well as collider experiments, and the indirect probe or constraint from the early Universe and the intensity frontier has expanded the WIMP paradigm beyond weak scale.

Quite recently, a tantalizing hint has been announced for the potential dark matter signals from the electron recoil events in the recoil energy, $E_R=1-10\,{\rm keV}$, from XENON1T experiment \cite{xenon}. 
The origin of the Xenon excess has been pondered over by particle physicists as seen from a lot of the already published articles on the arXiv for the last few days \cite{fumi,raidal,darkphoton,harigaya,rest}. 
A simple explanation with the solar axion or the neutrino magnetic dipole moment has been put forward from the XENON1T collaboration, but both cases are inconsistent with the star cooling constraints, because the electron coupling to the axion or the neutrino magnetic dipole moment required for the Xenon excess exceeds them by the order of magnitude \cite{xenon}.

In this article, we consider the possibility that exothermic dark matter transits from the heavy state to the light state in the event of the scattering with the electron. This possibility was already discussed in a different context in the literature with a motivation to explain the annual modulation signal at DAMA/LIBRA \cite{exothermic}. In this scenario, the down-scattering of dark matter makes the recoiled energy of the electron much larger than the one inferred from the elastic scattering between the non-relativistic dark matter in the standard halo model and the electron.
We discuss the details of the kinematics of exothermic dark matter and calculate the scattering cross section for the Xenon events by including the phase-space enhancement for inelastic scattering and the atomic excitation factor for a small momentum transfer between dark matter and electron.  We infer the required scattering cross section for dark matter as a function of dark matter mass at a fixed recoil energy near $E_R=2-3\,{\rm keV}$ up to the detector resolution.  We remark that there was a recent discussion on the monochromatic electron recoil spectrum in the case of $3\rightarrow 2$ inelastic scattering between dark matter and electron \cite{3to2}.

We also provide the model-independent discussion with a massive $Z'$ mediator on the effective model parameters explaining the Xenon excess and the dark matter relic density. We develop it for microscopic origins of the dark matter transition as well as the electron coupling, based on the $Z'$-portal and the vector--like lepton portal.

A similar idea has been discussed in Ref.~\cite{harigaya} while this article is being finalized. Our results agree with theirs and complement with the detailed dynamics of exothermic dark matter such as the phase-space enhancement, the constraints from dark matter relic density, and concrete microscopic models realizing the scenario.

\section{Exothermic dark matter and electron recoil}

In this section, we begin with the kinematics for exothermic dark matter in the case of down-scattering off the electron.
Then, we calculate the event rate for the electron recoil in the Xenon atoms by including the phase-space enhancement and the atomic excitation factor.

\subsection{Kinematics for exothermic dark matter}

We first consider the inelastic scattering between dark matter and electron, $\chi_1 e\rightarrow \chi_2 e$, where two dark fermions, $\chi_1$ and $\chi_2$, have a small mass difference, $\Delta m=m_{\chi_1}-m_{\chi_2}>0$, which is the exothermal condition for the transition in the dark matter states.
We assume that only $\chi_1$ or both $\chi_1$ and $\chi_2$ account for the observed relic density for dark matter.
The up-scattering process, $\chi_2 e\rightarrow \chi_1 e$, is also possible if kinematically allowed in the tail of the dark matter velocity distribution, for instance, at $v\sim 0.1$ for $\Delta m\sim {\rm keV}$. But, we regard the up-scattering process as being effectively forbidden for the standard halo model. 
Thus, we focus on the down-scattering process, $\chi_1 e\rightarrow \chi_2 e$, in the following discussion.

The electrons bound to Xenon atoms need the ionization energy to be excited to free electrons, and they
carry a nonzero velocity, $v_e\sim Z_{\rm eff} \alpha\sim 10^{-2}$, where $Z_{\rm eff}=1$ for outer shell electrons and it is larger for inner shells. Moreover, since electrons are bound to the atom, the entire atom recoils. In this case, the energy transferred to the electron can be obtained in terms of the energy lost in the dark matter and the nuclear recoil energy \cite{general}, as follows,
\bea
\Delta E_e &=& -\Delta E_{\rm DM} -\Delta E_N \nonumber \\
&=&\Delta m \bigg( 1-\frac{1}{2}\Big(\frac{m_{\chi_1}}{m_{\chi_2}} \Big) v^2 \bigg) + \frac{m_{\chi_1}}{m_{\chi_2}} \,{\vec q}\cdot {\vec v} -\frac{{\vec q}^2}{2\mu_{\chi_2N}} \nonumber \\
&=&\Delta m \bigg( 1-\frac{1}{2}\Big(\frac{m_{\chi_1}}{m_{\chi_2}} \Big) v^2 \bigg)-\frac{1}{2\mu_{\chi_2 N}}\bigg( {\vec q}-\frac{m_{\chi_1}}{m_{\chi_2}}\, \mu_{\chi_2 N}\, {\vec v} \bigg)^2 \nonumber \\
&\leq & \Delta m \bigg( 1-\frac{1}{2}\Big(\frac{m_{\chi_1}}{m_{\chi_2}} \Big) v^2 \bigg) + \frac{1}{2} \mu_{\chi_2 N} \Big(\frac{m_{\chi_1}}{m_{\chi_2}}\Big)^2 v^2
 \label{general}
\eea 
where $\mu_{\chi_2N}$ is the reduced mass for dark matter $\chi_2$ and the nucleus, $\mu_{\chi_2N}=m_{\chi_2} m_N/(m_{\chi_2} +m_N)$. Thus, the electron recoil energy is maximized if $ {\vec q}=\frac{m_{\chi_1}}{m_{\chi_2}}\, \mu_{\chi_2 N}\, {\vec v}$. Due to the bound electrons, the recoil energy of the ionized electron is given by $E_R=\Delta E_e-E_{Bi}$ where $E_{Bi}$ is the binding energy of the electron.
We note that there is a crucial difference from the elastic scattering, in that the upper bound on the electron energy is increased by the mass difference between dark matter components.

 For $\Delta m=0$, the typical momentum transfer in the elastic scattering between dark matter and electron is given by $q_{\rm typ}\sim \mu_{\chi e} v_{\rm rel}\sim m_e v_e\sim Z_{\rm eff} \,(4\,{\rm keV})$ for $m_\chi \gtrsim 1\,{\rm MeV}$.  But, for $\Delta m\neq 0$, the approximate momentum transfer given by  eq.(\ref{qapprox}) becomes almost independent of the velocity and $q^2\sim m_e \Delta m\gg q^2_{\rm typ}\sim m^2_e v^2_e$ for $\Delta m \gg m_e v^2_e\sim 0.05\,{\rm keV}$. 
  Then, we can also ignore the nuclear recoil energy in the above formula (\ref{general}), because $\Delta E_N=\frac{{\vec q}^2}{2\mu_{\chi_2 N}}\sim \frac{m_e}{m_{\chi_2}} (\Delta m) \ll \Delta m$ for $m_e\ll m_{\chi_2}<m_N$.

In the following discussion, we include a nonzero velocity of electron in the lab frame and ignore the nuclear recoil effects to show the salient effect of the mass splitting for the electron recoil energy. 
We take the electron and dark matter $\chi_1$ to have nonzero velocities  in the lab frame, $v_e$ and $v$, respectively, then their initial kinetic energies are given by $E_e=\frac{1}{2}m_e v^2_e$ and $E_0=\frac{1}{2} m_{\chi_1} v^2$.
From the energy conservation, the total energy after scattering satisfies the following relation,
\bea
m_{\chi_2}+\frac{p^2}{2M_2} + \frac{p^2_{\rm cm}}{2\mu_2} = m_{\chi_1}+E_0 +E_e  \label{energyconserv}
\eea
where $\mu_2$ is the reduced mass of the dark matter-electron system after scattering, given by
$\mu_2=m_e m_{\chi_2}/(m_e+ m_{\chi_2})$, and $M_2=m_e+m_{\chi_2}$, $p$ is the total 3-momentum and $p_{\rm cm}$ is the dark matter momentum in the center of mass frame after scattering.
Then, we can solve eq.~(\ref{energyconserv}) to get
\bea
p_{\rm cm} =\sqrt{2\mu_2 \Big(\Delta m +E_0 + E_e -\frac{p^2}{2M_2} \Big)} \label{pcm}
\eea
with $ \Delta m=m_{\chi_1} - m_{\chi_2}$.
On the other hand, we also get the total 3-momentum in terms of the initial kinetic energies, 
\bea
p^2=2m_{\chi_1} E_0 +2m_e E_e +4\cos\alpha\sqrt{m_e m_{\chi_1}E_0 E_e }  \label{tmom}
\eea
with $\alpha$ being the angle between electron and dark matter velocities. 

The electron recoil energy is given by the difference between the electron kinetic energies before and after scattering, 
\bea
\Delta E_e&=& \frac{1}{2m_e} \bigg( \frac{m_e}{M_2} {\vec p}-{\vec p}_{\rm cm}\bigg)^2- E_e \nonumber \\
&=&  \frac{1}{2m_e} \bigg(\frac{m^2_e}{M^2_2}\,p^2 +p^2_{\rm cm} -\frac{2m_e}{M_2}\,p\,p_{\rm cm} \cos\theta \bigg) - E_e
\eea
where $\theta$ is the scattering angle in the center of mass frame.
Then, using eq.~(\ref{pcm}), we obtain the exact expression for the electron recoil energy in terms of as
\bea
\Delta E_e&=&\frac{m_e}{2M^2_2}\, p^2 +\frac{\mu_2}{m_e} \Big(\Delta m+E_0 +E_e\Big)-\frac{\mu_2}{ 2m_eM_2}\,p^2 -E_e \nonumber \\
&& -\frac{p}{M_2}\,\cos\theta  \sqrt{2\mu_2(\Delta m +E_0+E_e)-\frac{\mu_2}{M_2}\,p^2}.\label{electron-recoil}
\eea  
On the other hand, the 3-momentum transfer $q$ is also given by
\bea
q^2=2m_e \Delta E_e +m^2_e v^2_e -2m_e v_e\cos\psi \sqrt{m_e \Delta E_e} \label{momentum-transf}
\eea
where $\psi$ is the scattering angle of the electron in the lab frame. Here, we note that $q^2=2m_e \Delta E_e$ only for the electron at rest, which is the standard relation between the recoil energy and the electron, when the target electron is at rest.

In the following subsections, we divide our discussion on the electron recoil energy, depending on the masses of dark matter components.

\subsubsection{Heavy dark matter}

First, we take the limit with $m_{\chi_1} E_0\gtrsim m_e E_e$, which is the case for $m_{\chi_1}\gtrsim m_e (v_e/v)\sim 10\,{\rm MeV}$, i.e. heavy dark matter. In the next section, our detailed discussion on the Xenon electron events will be based on this case. Then, we can approximate the total 3-momentum to be $p^2\simeq 2m_{\chi_1} E_0$ from eq.~(\ref{tmom}). As a result,  from eq.~(\ref{electron-recoil}), we obtain the approximate electron recoil energy as follows,
\bea
\Delta E_e \simeq \frac{\mu_2 \Delta m}{m_e} \bigg(1-\frac{E_0}{M_2} \bigg)-\frac{m_e}{M_2}\,E_e +\frac{\mu^2_2 E_0 }{m_e m_{\chi_2}}\bigg[\bigg(1+\frac{m_{\chi_1}}{m_{\chi_2}} \bigg)-2\sqrt{\kappa} \cos\theta\bigg] \label{recoil-approx}
\eea
with
\bea
\kappa &\equiv& \frac{m_{\chi_1}}{m_{\chi_2}}\bigg(1+\frac{m_{\chi_1} E_e}{\mu_2 E_0} \bigg)+\frac{m_{\chi_1} \Delta m}{\mu_2 E_0} \Big(1-\frac{E_0}{M_2} \Big)
\nonumber \\
&=& \frac{m_{\chi_1}}{m_{\chi_2}}\bigg(1+\frac{m_e v^2_e}{\mu_2 v^2} \bigg)+\frac{2\Delta m}{\mu_2 v^2} \bigg(1-\frac{\mu_2 v^2}{2m_e}\frac{m_{\chi_1}}{m_{\chi_2}} \bigg). \label{kappa}
\eea

In the case for heavy dark matter, taking the limits for $\Delta m\ll m_e\ll m_{\chi_1}$ and $\Delta m\gg m_e v^2_e\sim 50\,{\rm eV}$, we can approximate eq.~(\ref{kappa}) as
\bea
\kappa&\simeq& \frac{m_e v^2_e+2\Delta m}{m_e v^2} \nonumber \\
&\simeq& \frac{\Delta m}{ \frac{1}{2} m_e v^2}\simeq 2.2\times 10^4 \bigg(\frac{220\,{\rm km/s}}{v}\bigg)^2 \bigg(\frac{\Delta m}{3\,{\rm keV}}\bigg).
\eea
Indeed, choosing $\Delta m\simeq 3\,{\rm keV}$ from Xenon electron recoil energy and $v\simeq 220\,{\rm km} $ from the averaged velocity of dark matter in the standard halo model, we obtain  $\kappa\gg 1$.
In this case, from eq.~(\ref{recoil-approx}), the electron recoil energy becomes simplified to
\bea
\Delta E_e &\simeq&  \Delta m \bigg( 1-\frac{1}{2}\Big(\frac{m_{\chi_1}}{m_{\chi_2}} \Big) v^2 \bigg) -\frac{m^2_e v^2_e}{2m_{\chi_1}}+ m_e v^2 \Big(1-\sqrt{\kappa} \cos\theta \Big)  \nonumber \\
&\simeq&  \Delta m  \bigg(1-\frac{2}{\sqrt{\kappa}}\cos\theta \bigg). \label{ERapprox} 
\eea
On the other hand, from eq.~(\ref{momentum-transf}), for $\Delta E_e\simeq \Delta m\gg m_e v^2_e$, we can also get the approximate result for the momentum transfer as
\bea
q^2 \simeq 2m_e \Delta E_e
\simeq 2m_e \Delta m \bigg(1-\frac{2}{\sqrt{\kappa}}\, \cos\theta \bigg). \label{qapprox}
\eea
Therefore,  either the recoil energy or the momentum transfer depend little on either the dark matter and electron velocities or the scattering angle. As a result, we get a tiny momentum transfer fixed by the mass difference, for which the atomic excitation factor becomes important  \cite{ionization,excitation,raidal,harigaya}, as will be discussed later. 

For $\Delta m\ll m_e\ll m_{\chi_1}$ but with $\kappa\simeq 1$, we have the following approximate results,
\bea
\Delta E_e &\simeq & \Delta m +m_e v^2 (1-\cos\theta), \\
q^2 &\simeq & 2m^2_e v^2 (1-\cos\theta).
\eea
In this case, the recoil energy is bounded by $2m_e v^2\simeq 5.5\times 10^{-4}\,{\rm keV}\gg \Delta m$ for $v\sim 220\,{\rm km/s}$, which is too small to account for the Xenon experiment. 
Thus, we focus on the regime with $\kappa\gg 1$ in the following discussion, which is sufficient for explaining the Xenon excess.

\subsubsection{Light dark matter}

In the opposite limit with $m_{\chi_1} E_0\lesssim m_e E_e$, that is,  $m_{\chi_1}\lesssim m_e (v_e/v)\sim 10\,{\rm MeV}$, that is, light dark matter, we can approximate the total momentum as $p^2\simeq 2m_e E_e$. In this case,  from eq.~(\ref{electron-recoil}), we get the approximate recoil energy as follows,
\bea
\Delta E_e \simeq \frac{\mu_2}{m_e} \,( \Delta m+E_0 )-\frac{2\mu_2}{M_2}\,E_e \Big(1+\sqrt{\tilde\kappa} \cos\theta\Big) \label{recoil-approx2}
\eea
with
\bea
{\tilde\kappa} &\equiv& 1+\frac{m_e}{\mu_2 E_e}\, (\Delta m+E_0)
\nonumber \\
&=&1+\frac{1}{\mu_2 v^2_e}\, (2\Delta m+m_{\chi_1} v^2) . 
\label{kappat}
\eea

In the case for light dark matter, taking the limits for $\Delta m\ll m_e\ll m_{\chi_1}$ and $m_{\chi_1} v^2\lesssim 100\,{\rm eV}\ll \Delta m$,
we can approximate eq.~(\ref{kappat}) as
\bea
{\tilde\kappa} \simeq  \frac{\Delta m}{\frac{1}{2}m_e v^2_e}\simeq 10^2\bigg(\frac{10^{-2}c}{v_e}\bigg)^2 \bigg(\frac{\Delta m}{3\,{\rm keV}} \bigg). 
\eea
Then, from eq.~(\ref{recoil-approx2}), the electron recoil energy becomes simplified to
\bea
\Delta E_e &\simeq & \Delta m \Big(1+\frac{m_{\chi_1}v^2}{2\Delta m} \Big) - \frac{m^2_e v^2_e}{m_{\chi_1}} \Big(1+\sqrt{\tilde\kappa} \cos\theta \Big) \nonumber \\
&\simeq & \Delta m \bigg(1-\frac{2m_e}{m_{\chi_1}} \, \frac{1}{\sqrt{{\tilde\kappa}}}\,\cos\theta\bigg). \label{ERapprox2}
\eea
On the other hand, from eq.~(\ref{momentum-transf}), for $\Delta E_e\simeq \Delta m\gg m_e v^2_e$, we can also get the approximate result for the momentum transfer as
\bea
q^2 \simeq 2m_e \Delta E_e
\simeq 2m_e \Delta m \bigg(1-\frac{2m_e}{m_{\chi_1}} \, \frac{1}{\sqrt{{\tilde\kappa}}}\,\cos\theta\bigg). \label{qapprox2}
\eea
Then, as in the case for heavy dark matter, the electron recoil energy depends little on either the dark matter and electron velocities or the scattering angle, but it is determined dominantly by the mass difference.

\subsection{The event rate for electron recoil}

We begin with the general expression for the event rate per target mass \cite{DD}, given by
\bea
dR = \frac{\rho_{\chi_1} v}{m_{\chi_1} m_T}\, d\sigma\, f_1(v )dv \label{Rate}
\eea
where $m_T$ is the target nucleus mass and $f_1(v)=\frac{4v^2}{v^3_0 \sqrt{\pi}} \, e^{-v^2/v^2_0}$ with $v_0=220\,{\rm km/s}$   for the Maxwellian velocity distribution of dark matter and $\int^\infty_0 f_1(v) dv=1$, and  $\rho_{\chi_1}$ is the local energy density of dark matter,  which is given by $\rho_{\chi_1}=0.4\,{\rm GeV/cm^3}$ if $\chi_1$ occupies the full dark matter.

We note that the differential scattering cross section for the inelastic scattering is given by
\bea
\frac{d\sigma}{d E_R} =\frac{2m_e {\bar\sigma}_e}{q^2_+-q^2_-} \int^{q_+}_{q_-} a^2_0 \,  q'\,  dq'\, K(E_R,q')\, P^2(v), \label{dsigma}
\eea
where ${\bar\sigma}_e$ is the total cross section in the limit of elastic scattering for the fixed momentum transfer at $q=1/a_0$, with $a_0=\frac{1}{\alpha m_e}$ being the Bohr radius, $K(E_R,q')$ is the atomic enhancement factor, and  $P^2(v)$ is the phase space factor, which is unity for the elastic scattering. 
Then, from eqs.~(\ref{Rate}) and (\ref{dsigma}), we obtain the differential event rate per target mass as
\bea
\frac{dR}{dE_R}= \frac{2 m_e {\bar\sigma}_e \rho_{\chi_1}}{m_{\chi_1} m_T} \,K_{\rm int}(E_R) \int^\infty_{v_{\rm min}}\, \frac{v P^2(v)}{q^2_+-q^2_-}\, f_1(v)\, dv \label{diffR}
\label{dR}
\eea
where the integrated atomic enhancement factor is given by
\bea 
K_{\rm int}(E_R) = \int^{q_+}_{q_-} a^2_0\,  q'\,  dq'\, K(E_R,q'),
\eea 
and $v_{\rm min}$ is the minimum velocity of dark matter required for a given recoil energy $E_R$.
We note that the total recoil energy is also deposited significantly near $E_R\sim {\rm keV}$ to ionize the electrons bound to the Xenon atoms, and the atomic excitation factor can be important for a small momentum transfer \cite{ionization,excitation,raidal,harigaya}. 
As a result, we get  the event rate per detector as
\bea
R_D=M_T  \int_{E_T}^\infty \frac{dR}{ dE_R} \, dE_R \label{totalR}
\eea
where $E_T$ is the detector threshold energy and $M_T$ is the fiducial mass of the detector, given by $M_T\simeq 4.2\times 10^{27}(M_T/{\rm tonne}) m_T$ for Xenon.

Now we apply the general result in eq.~(\ref{dR}) for the case with down-scattering dark matter.
Assuming that dark matter mass $m_{\chi_1}$ is greater than $10\,{\rm MeV}$, we can use our results for heavy dark matter in Section 2.1.1.
Then, we take $\kappa\gg 1$, for which the recoil energy is appreciable.  In this case, we obtain the phase space factor $P^2(v)$ in eq.~(\ref{diffR}) as
\bea
P^2(v)\simeq \sqrt{1+\frac{2\Delta m}{\mu_1 v^2}}\simeq  \sqrt{\frac{2\Delta m}{m_e}}\, \frac{1}{v}.  \label{enhance}
\eea
Then, using eq.~(\ref{dR}) with $q^2_\pm\simeq 2m_e \Delta m \big(1\pm \frac{2}{\sqrt{\kappa}} \big)$ from eq.~(\ref{qapprox}),
we obtain the differential event rate  for $E_-<E_R<E_+$ with $E_\pm =\Delta m \big(1\pm \frac{2}{\sqrt{\kappa}} \big)$ as
\bea
\frac{dR}{ dE_R} &\simeq & \frac{{\bar\sigma}_e \rho_{\chi_1}}{2 m_e m_{\chi_1} m_T} \, K_{\rm int}(E_R) \int_0^{v_{\rm max}}  \frac{f_1(v)}{v} dv\, \theta(E_R-E_-)\theta(E_+-E_R) \label{diffR}
\eea
where $v_{\rm min}=0$, $v_{\rm max}=\sqrt{\frac{2\Delta m}{m_e}}$ at $\kappa= 1$. 
For $E_+-E_-\ll E_{\pm}$, we can approximate $\theta(E_R-E_-)\theta(E_+-E_R)\simeq (E_+-E_-) \delta(E_R-\Delta m)$. 
Therefore, we can rewrite  eq.~(\ref{diffR}) as
\bea
\frac{dR}{ dE_R} &\simeq &\Big( \frac{2\Delta m}{m_e}\Big)^{1/2} \frac{  {\bar\sigma}_e \rho_{\chi_1}}{m_{\chi_1} m_T}\, K_{\rm int}(E_R) \delta(E_R-\Delta m)\int_0^{v_{\rm max}}   f_1(v) dv \label{diffR2}
\eea
We note that for $\Delta m\gg 1.3\times 10^{-4}\,{\rm keV}$, we have $v_{\rm max}\gg v_0$, resulting in $\int_0^{v_{\rm max}}   f_1(v) dv\simeq 1$. Therefore, we find that there is no Boltzmann suppression due to an enhancement factor $P^2(v)$ in eq.~(\ref{enhance}), as compared to the case with elastic scattering.

Consequently, from eq.~(\ref{totalR}) with eq.~(\ref{diffR2}), we get the total  event rate per Xenon detector with $m_T=m_{Xe}$ as
\bea
R_D &\simeq& \bigg( \frac{ M_T\, {\bar\sigma}_e\,  \rho_{\chi_1}}{m_{\chi_1} m_T}\bigg) \bigg( \frac{2\Delta m}{m_e}\bigg)^{1/2}K_{\rm int}(\Delta m)   \nonumber \\
&\simeq &50 \bigg(\frac{M_T}{\rm tonne-yrs}\bigg) \bigg(\frac{K_{\rm int}(\Delta m) }{2.6} \bigg) \bigg(\frac{\rho_{\chi_
1}}{0.4\,{\rm GeV\, cm^{-3}}}\bigg) \nonumber \\
&&\times \bigg(\frac{{\bar\sigma}_e/m_{\chi_1}}{1.2\times 10^{-43}\,{\rm cm}^2/{\rm GeV}} \bigg) \bigg(\frac{\Delta m}{2.5\,{\rm keV}}\bigg)^{1/2} \label{total}
\eea
where we has used the normalization for the integrated atomic excitation factor at $E_R\simeq 2\,{\rm keV}$ and for the momentum transfer peaked at $q\simeq 50\,{\rm keV}$.

For comparison to the experimental data, the mono-energetic event rate can be convoluted with the detector resolution by
\bea
\frac{dR_D}{dE_R} = \frac{R_D}{\sqrt{2\pi} \sigma} \, e^{-(E_R-\Delta m)^2/(2\sigma^2)}\,\alpha(E)
\eea
where $\sigma$ is the detector resolution, which varies between $20\%$ at $E=2\,{\rm keV}$ and $6\%$ at $E=30\,{\rm keV}$, and $\alpha(E)$ is the signal efficiency \cite{xenon}. 
For $E_R=2-10\,{\rm keV}$, the signal efficiency is given by $\alpha(E)\sim 0.7-0.9$ \cite{xenon}.

\begin{figure}[tbp]
  \centering
\includegraphics[width=.60\textwidth]{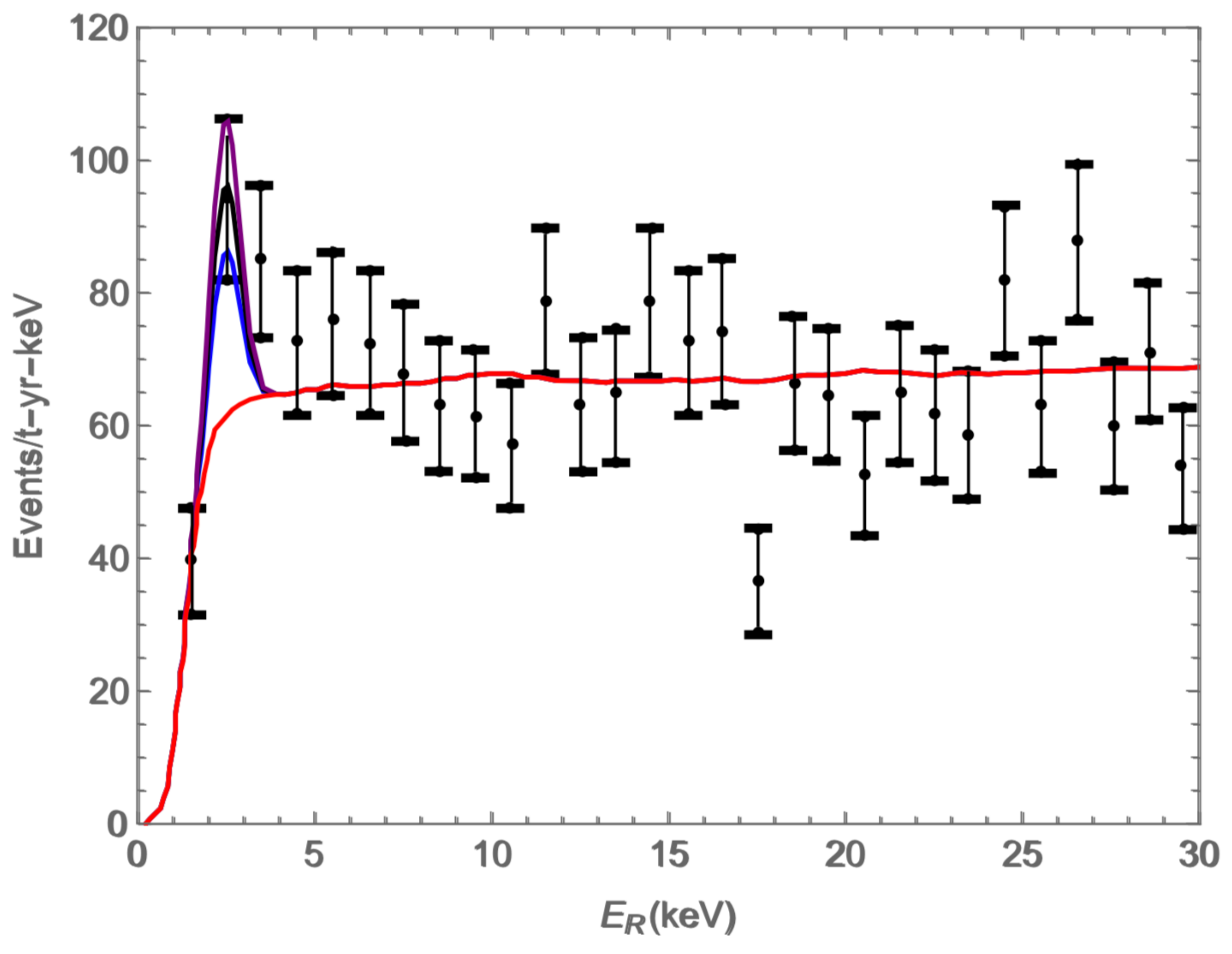}
  \caption{ The event rate for electron recoil as a function of recoil energy $E_R$ in keV. We have taken $\Delta m=2.5\,{\rm keV}$ and ${\bar\sigma}_e/m_{\chi_1}=1.0, 1.4, 1.8\times 10^{-43}\,{\rm cm}^2$ from bottom to top lines (purple, black and blue. The red line is the background model used by Xenon experiment \cite{xenon}.}
  \label{RD}
\end{figure}

The XENON1T excess is most significant from the electrons at $E_R=2-3\,{\rm keV}$ with the detector resolution being about $\sigma=0.4\,{\rm keV}$, so we take the recoil energy of the mono-energetic electron in our model to be $E_R\simeq \Delta m\simeq 2.5\,{\rm keV}$. Moreover, from $\alpha(E)\simeq 0.8$ at $E_R\simeq 2.5\,{\rm keV}$, we need to rescale the total event rate per detector in eq.~(\ref{total}) by a factor $0.8$.  Therefore, taking into account the total exposure in XENON1T for SR1, which is $0.65$ tonne-yrs \cite{xenon}, and for $\rho_{\chi_1}=0.4\,{\rm GeV\, cm^{-3}}$,  we can get about 50 events near $E_R=2-3\,{\rm keV}$ for the XENON1T electron recoil events for ${\bar\sigma}_e/m_{\chi_1}\simeq 3.4\times 10^{-43}\,{\rm cm^2}/{\rm GeV}$. 

In Fig.~\ref{RD}, we depict the expected event rate for the signal $+$ background in units of events per $\rm t-yr-keV$ as a function of the electron recoil energy in ${\rm keV}$ in our model with exothermic dark matter.  
The expected event rates are shown in blue lines for $\Delta m=2.5\,{\rm keV}$ and ${\bar\sigma}_e/m_{\chi_1}=1.0, 1.4, 1.8\times 10^{-43}\,{\rm cm}^2$, from bottom to top, and the red line corresponds to the background model used by Xenon experiment \cite{xenon}. The values of the mass difference and ${\bar\sigma}_e/m_{\chi_1}$ can be varied for the global fit with the observed electron recoil spectrum, but the monochromatic shape of the signal from exothermic dark matter above the flat background remains the same up to the detector resolution. 
We can compare our case to the existing global fit of the XENON electron recoil spectrum for the case with axion-like dark matter whose mass is fixed to $m_a=2.3\pm 0.2$ keV with $3.0\sigma$ local significance \cite{xenon}.

\section{The effective theory for exothermic dark matter}

We continue to discuss the effective theory for exothermic dark matter in the presence of a massive $Z'$ mediator and constrain the parameter space for the $Z'$ couplings and  the mass parameters from the Xenon excess.
For completeness, we also provide the formulas for dark matter annihilation cross sections  in the effective theory and comment on the compatibility of the Xenon excess with the correct relic density.

\subsection{The effective interactions and the Xenon excess}

We consider two Majorana dark matter fermions, $\chi_1$ and $\chi_2$, with different masses, $m_{\chi_1}>m_{\chi_2}$, and a massive dark gauge boson $Z'$ with mass $m_{Z'}$.
We take the effective Lagrangian with $Z'$ couplings to dark fermions, electron and electron neutrino, in the following form,
\bea
{\cal L}_{\rm eff} &=&  \Big(g_{Z'}  Z'_\mu\, {\bar \chi}_2\gamma^\mu (v_\chi+a_\chi\gamma^5 ) \chi_1 +{\rm h.c.} \Big)+  g_{Z'}  Z'_\mu\, {\bar e}(v_e + a_e \gamma^5) e  \nonumber \\
&&+ g_{Z'}  Z'_\mu\, {\bar \nu}\gamma^\mu (v_\nu+a_\nu \gamma^5 ) \nu
 \label{eft}
\eea
where $v_i, a_i$ with $i=\chi, e, \nu$ are constant parameters.
In the next subsection, we will show a microscopic model for the above effective interactions. 
For $\Delta m<m_{Z'}$  and $\Delta m< 2m_e$, there is no tree-level decay process for the dark matter fermion $\chi_1$.

For $a_e\neq 0$, the effective vertex interaction for one $Z'$ and two photons, $Z'_\mu(q)-A_\lambda(q_1)-A_\sigma(q_2)$, with $q=q_1+q_2$, is induced by the electron loops, taking the following form for $q^2\ll 4m^2_e$ \cite{tait},
\bea
{\Gamma}^{\mu\lambda\sigma}(q,q_1,q_2) \simeq \epsilon^{q_1\lambda\sigma\mu}\,\cdot\frac{a_e e^2  g_{Z'}}{4\pi^2}\, \bigg( 1+ \frac{q^2}{12m^2_e} \bigg).
\eea
Then, the heavy state of dark matter can decay into the lighter state of dark matter and two photons, with the decay rate given by
\bea
\Gamma(\chi_1\rightarrow \chi_2 \gamma\gamma) \simeq \frac{a^2_e( v^2_\chi+a^2_\chi) e^4 g^2_{Z'}}{2560\pi^7}\frac{(\Delta m)^5}{ m^4_{Z'}}.
\eea
Therefore, in this case, the lifetime of the dark fermion $\chi_1$ is much longer than the age of the Universe for perturbative effective couplings. However, the diffuse X-ray background \cite{xray} puts the bound on the lifetime of the dark fermion $\chi_1$ to $\tau_{\chi_1}>10^{24}\,{\rm sec}$,  which gives rise to
\bea
|a_e| g_{Z'} \sqrt{v^2_\chi+a^2_\chi}< 2.5\times 10^{-6}\, \bigg(\frac{2.5\,{\rm keV}}{\Delta m} \bigg)^{5/2} \bigg( \frac{m_{Z'}}{1\,{\rm GeV}}\bigg)^2.
\eea
As a result, we need a small axial vector coupling to the electron to satisfy the X-ray bounds for exothermic dark matter and light $Z'$ mediator.
As we will discuss in the next section, some microscopic models with vector-like leptons can induce a suppressed axial vector coupling to the electron.

Moreover, there is another loop process for the three-photon decay channel, $\chi_1\rightarrow \chi_2+3\gamma$, but the corresponding decay rate is highly suppressed by $\Gamma\propto (\Delta m)^{13}/(m^4_{Z'}m^8_e)$ \cite{harigaya}, thus being consistent with X-ray bounds \cite{xray}.   

If the neutrino couplings to $Z'$ are nonzero,  the dark matter fermion $\chi_1$ would decay into a neutrino pair  via the off-shell $Z'$ gauge boson, which is bounded by the lifetime of dark matter for explaining the XENON1T electron recoil excess. 
The decay rate of the dark fermion $\chi_1$ is given \cite{cbpark} by
\bea
\Gamma(\chi_1\rightarrow \chi_2 \nu {\bar\nu}) \simeq \frac{N_\nu G^{\prime 2}_F (\Delta m)^5}{30\pi^3} \, (v^2_\chi + 3 a^2_\chi) \, (v_\nu^2+a_\nu^2)
\eea
where $N_\nu$ is the number of neutrinos coupled to $Z'$, and  $G'_F \equiv g^2_{Z'}/(\sqrt{2} m^2_{Z'})$.
Then, for $\Delta m=2.5\,{\rm keV}$ and $v_\chi=-a_\chi=\frac{1}{2}$, the lifetime of the dark fermion $\chi_1$ is longer than the age of the Universe, as far as 
\bea
G^{\prime }_F  \sqrt{N_\nu(v_\nu^2+a_\nu^2)} <2.4 \times 10^{-6}\, {\rm GeV}^{-2}. \label{lifetime}
\eea
Nonetheless, neutrino experiments such as Super-Kamiokande \cite{SK} would constrain the lifetime of dark matter to $\tau_{\chi_1}>10^{24}\,{\rm sec}$ for $\Delta m>0.1\,{\rm MeV}$ \cite{ibarra}. But, for $\Delta m=2.5\,{\rm keV}$, the energy of the produced neutrinos is below those of solar neutrinos, so there is no current bound from neutrino experiments. 
As will be shown in the next section, the effective neutrino couplings are induced in the case of $Z'$ portal with a gauge kinetic mixing, but there is no bound on them other than the lifetime bound from the age of the Universe.

For $m_e, m_{\chi_1}, m_{Z'}\gg q\sim m_e \Delta m$, $m_e\ll m_{\chi_1}$, and $\Delta m\ll m_{\chi_1}$,  the total scattering cross section for $\chi_1 e\rightarrow \chi_2 e$, up to the phase space factor $P^2(v)$ in eq.~(\ref{enhance}), is given by
\bea
{\bar\sigma}_{e}&\simeq& \frac{v^2_\chi v^2_e  g^4_{Z'}\mu^2_1}{\pi m^4_{Z'}} \nonumber \\
&\simeq & \bigg( \frac{v_\chi g_{Z'}}{0.6}\bigg)^2\bigg(\frac{v_e g_{Z'} }{10^{-4}e} \bigg)^2\bigg(\frac{1\,{\rm GeV}}{m_{Z'}} \bigg)^4\Big( \frac{\mu_1}{m_e}\Big)^2 \, \times 10^{-44}\,{\rm cm}^2\label{scatt}
\eea
where $e$ is the electromagnetic coupling.
Thus, we need to have nonzero vector couplings to both dark matter and electron for the scattering cross section without velocity suppression. 
In order to explain the XENON1T electron recoil events near $E_R=2-3\,{\rm keV}$ in our model, we take $\Delta m\simeq 2.5\,{\rm keV}$ and the required scattering cross section gives rise to the following useful formula,
\bea
\bigg( \frac{v_\chi g_{Z'}}{0.6}\bigg)^2\bigg(\frac{v_e g_{Z'} }{10^{-4}e} \bigg)^2\bigg(\frac{1\,{\rm GeV}}{m_{Z'}} \bigg)^4 \bigg(\frac{0.3\,{\rm GeV}}{m_{\chi_1}}\bigg)\bigg(\frac{\Omega_{\rm DM}}{\Omega_{\chi_1}}\bigg)\simeq 1 \label{xenon}
\eea
where $\Omega_{\rm DM}$ is the observed total abundance of dark matter and $\Omega_{\chi_1}$  is the abundance of the dark fermion $\chi_1$.
Therefore, light dark matter and $Z'$ mediator are favored by the explanation of the Xenon excess with exothermic dark matter.
As we scale up the $Z'$ gauge coupling, we can take a larger value of $m^4_{Z'}  m_{\chi_1}$ in order to maintain the number of the electron recoil events.

We remark on the choices of the $Z'$ couplings in view of the Xenon electron excess. 
First, we took the $Z'$ couplings to be consistent with the dilepton bounds from BaBar, $|v_e| g_{Z'}\lesssim 10^{-4}e$ for $0.02\,{\rm GeV}<m_{Z'}<10.2\,{\rm GeV}$ \cite{babar}, or the bound from mono-photon $+$ MET from BaBar \cite{babarinv}, $|v_e| g_{Z'}\lesssim (4\times 10^{-4}-10^{-3})\,e$ for $m_{Z'}<8\,{\rm GeV}$. 
There are other bounds from beam dump experiments \cite{beamdump} that limit $|v_e| g_{Z'}$ at the level of $10^{-3}$ or stronger  for $m_{Z'}\lesssim 0.1\,{\rm GeV}$ than in BaBar experiment. In the parameter space of our interest, we focus on $m_{Z'}\gtrsim 0.1\,{\rm GeV}$, for which the BaBar bounds are most stringent at present. Thus, we take into account only the BaBar  bounds in the later discussion on phenomenological constants in the next subsection and the next section on microscopic models.

\subsection{Dark matter annihilation and relic density}

Dark matter fermions $\chi_1$ and $\chi_2$ can co-annihilate into a pair of electrons as well as into a pair of $Z'$ gauge bosons if kinematically allowed.
Then, taking $\Delta m\ll m_{\chi_1}$ and ignoring the lepton masses, the total annihilation cross section for $n_{\rm DM}=n_{\chi_1}+n_{\chi_2}$ with $\chi_1{\bar \chi}_2\rightarrow e{\bar e}, \nu{\bar \nu}$ and $\chi_1{\bar \chi}_1,\chi_2{\bar \chi}_2\rightarrow Z'Z'$ is given by
\bea
\langle\sigma v\rangle=\frac{1}{2}\langle\sigma v\rangle_{\chi_1{\bar \chi}_2\rightarrow e{\bar e}, \nu{\bar \nu} }+\frac{1}{2}\langle\sigma v\rangle_{\chi_1{\bar \chi}_2\rightarrow Z'Z' }  \label{totalann}
\eea
with
\bea
\langle\sigma v\rangle_{\chi_1{\bar \chi}_2\rightarrow e{\bar e},\nu {\bar \nu} } &=&  \frac{g^4_{Z'} v^2_\chi}{\pi}\, \Big[v^2_e+a^2_e+N_\nu (v^2_\nu+a^2_\nu)\Big] \, \frac{m^2_{\chi_1}}{(m^2_{Z'}-4m^2_{\chi_1})^2 +\Gamma^2_{Z'} m^2_{Z'}}, \\
\langle\sigma v\rangle_{\chi_1{\bar \chi}_1,\chi_2{\bar \chi}_2\rightarrow Z'Z' }&=&  \frac{ g^4_{Z'} }{4\pi }\,\bigg[v^4_\chi +a^4_\chi+2v^2_\chi a^2_\chi \Big(4\frac{m^2_{\chi_1}}{m^2_{Z'}}-3 \Big) \bigg]\frac{m^2_{\chi_1}}{(m^2_{Z'}-2m^2_{\chi_1})^2}  \bigg(1-\frac{m^2_{Z'}}{m^2_{\chi_1}} \bigg)^{3/2}. 
\eea
We note that the contributions coming from $a_\chi$ to the annihilation cross section are $p$-wave suppressed.
Since we need $v_\chi\neq 0$ for explaining the Xenon electron excess, the $p$-wave annihilations are sub-dominant.

For light dark matter with sub-GeV mass, once $\chi_1{\bar \chi}_1,\chi_2{\bar \chi}_2\rightarrow Z'Z'$ is open, the resultant annihilation cross section would be too large for a sizable $g_{Z'}$ to account for the correct relic density.
Thus, in this case, we can take $m_{\chi_1}<m_{Z'}$ such that the annihilation of the dark matter fermion $\chi_1$ into a pair of $Z'$ is forbidden at zero temperature, but it is open in the tail of the Boltzmann distribution at a finite temperature during freeze-out \cite{fdm}.
Then, from the detailed balance condition for the forbidden channels, the effective annihilation cross section for the forbidden channels, $\chi_1{\bar \chi}_1, \chi_2{\bar \chi}_2\rightarrow Z'Z'$ becomes
\bea
\langle\sigma v\rangle_{\chi_1{\bar \chi}_1,\chi_2{\bar \chi}_2\rightarrow Z'Z' }=\frac{ (n^{\rm eq}_{Z'})^2}{n^{\rm eq}_{\chi_1} n^{\rm eq}_{\chi_2}}\, \langle\sigma v\rangle_{Z'Z'\rightarrow\chi_1{\bar \chi}_1,\chi_2{\bar \chi}_2}
\eea
with
\bea
\langle\sigma v\rangle_{Z'Z'\rightarrow\chi_1{\bar \chi}_1,\chi_2{\bar \chi}_2}&=&\frac{4 g^4_{Z'}}{9\pi m^2_{Z'}}\,\bigg[v^4_\chi +a^4_\chi \Big(\frac{m^2_{Z'}+2m^2_{\chi_1}}{m^2_{Z'}+m^2_{\chi_1}}\Big)+2v^2_\chi a^2_\chi\Big( \frac{3m^2_{Z'}-2m^2_{\chi_1}}{m^2_{Z'}+m^2_{\chi_1}} \Big) \bigg] \nonumber \\
&&\quad \times \bigg( 1+\frac{m^2_{\chi_1}}{m^2_{Z'}}\bigg)\bigg(1- \frac{m^2_{\chi_1}}{m^2_{Z'}} \bigg)^{3/2}.
\eea
Here, for $\Delta m\ll m_{\chi_1}$,  the Boltzmann suppression factor can be approximated to
\bea
\frac{ (n^{\rm eq}_{Z'})^2}{n^{\rm eq}_{\chi_1} n^{\rm eq}_{\chi_2}}\simeq \bigg(\frac{n^{\rm eq}_{Z'}}{n^{\rm eq}_{\chi_1}} \bigg)^2\simeq \frac{9}{4} \, \Big(\frac{m_{Z'}}{m_{\chi_1}} \Big)^3\, e^{-2(m_{Z'}-m_{\chi_1})/T}.
\eea
The forbidden channels are important for obtaining the correct relic density for light dark matter, because the strong annihilation cross section can be compensated by the Boltzmann suppression factor \cite{fdm}. 

Furthermore, for $m_{\chi_1}>m_{\chi_2}$, we have additional annihilation channels active even at zero temperature, $\chi_1 {\chi}_1\rightarrow \chi_2\chi_2$, $\chi_1{\bar\chi}_1\rightarrow \chi_2{\bar\chi}_2$, and their complex conjugates. Then, the additional annihilation cross sections for $\chi_1$ are given by
\bea
\langle\sigma v\rangle_{\chi_1\chi_1\rightarrow \chi_2\chi_2} &=&\frac{\sqrt{2}}{16\pi} \frac{g^4_{Z'} m^2_{\chi_1}}{m^4_{Z'}}\,\sqrt{\frac{\Delta m}{m_{\chi_1}}}\, (v^4_\chi+6a^2_\chi v^2_\chi+9 a^4_\chi), \\
 \langle\sigma v\rangle_{\chi_1{\bar\chi}_1\rightarrow \chi_2{\bar\chi}_2}&=&\frac{\sqrt{2}}{8\pi} \frac{g^4_{Z'} m^2_{\chi_1}}{m^4_{Z'}}\,\sqrt{\frac{\Delta m}{m_{\chi_1}}}\, (v^4_\chi+3 a^4_\chi).
\eea 
For $T\gtrsim \Delta m=2.5\,{\rm keV}$, the inverse annihilation processes, $\chi_2 {\chi}_2\rightarrow \chi_1\chi_1$, $\chi_2{\bar\chi}_2\rightarrow \chi_1{\bar\chi}_1$, and their complex conjugates, occur as often as the above annihilation, so the additional annihilation processes have no impact on the total dark matter density at the time of freeze-out of the annihilation processes, given in eq.~(\ref{totalann}).
However, the $\chi_1$ component could keep annihilating and its abundance would be Boltzmann suppressed by $e^{-\Delta/T_{\chi_1}}$ with $T_{\chi_1}=T^2/T_{\rm kd}$ where $T_{\rm kd}$ is the kinetic decoupling temperature, which is the smaller of the decoupling temperature of dark matter and the electron decoupling temperature $\sim 1\,{\rm MeV}$.
Then, in order to avoid the Boltzmann suppression for the relic abundance of the heavier component, we need to impose
\bea
T_{\chi_1}\gtrsim \Delta m, \label{latedec}
\eea
which corresponds to the bound on the radiation temperature,
\bea
T\gtrsim \sqrt{T_{\rm kd} \Delta m}. 
\eea
The decoupling temperature for the $2\to 2$ annihilation processes is determined by
\bea
n_{\chi_1}\, {\rm max} \Big(\langle\sigma v\rangle_{\chi_1\chi_1\rightarrow \chi_2\chi_2}, \langle\sigma v\rangle_{\chi_1{\bar\chi}_1\rightarrow \chi_2{\bar\chi}_2}\Big)=H.
\eea
Thus, imposing the above equation with eq.~(\ref{latedec}) and  for $|v_\chi|=|a_\chi|$,
we find the upper limit on the $Z'$ gauge coupling as follows,
\bea
\frac{ |v_\chi| g_{Z'} m_{\chi_1}}{m_{Z'}}\lesssim 0.035\bigg(\frac{\Omega_{\rm DM}/2} {\Omega_{\chi_1}}\bigg)^{1/4}\bigg(\frac{m_{\chi_1}}{100\,{\rm MeV}} \bigg)^{1/2} \bigg(\frac{m_{\chi_1}/\Delta m}{4\times 10^4}\bigg)^{3/8} \bigg(\frac{\Delta m/(2.5\,{\rm keV})}{T_{\rm kd}/(10\,{\rm MeV})}\bigg)^{1/8}. \label{latedec2}
\eea
Therefore, we need to impose the above condition that the annihilation processes, $\chi_1 {\chi}_1\rightarrow \chi_2\chi_2$, $\chi_1{\bar\chi}_1\rightarrow \chi_2{\bar\chi}_2$,  decouple sufficiently early at $T_{\chi_1}\gtrsim\Delta m$, in the following discussion on the relic density.

\begin{figure}[tbp]
  \centering
\includegraphics[width=.40\textwidth]{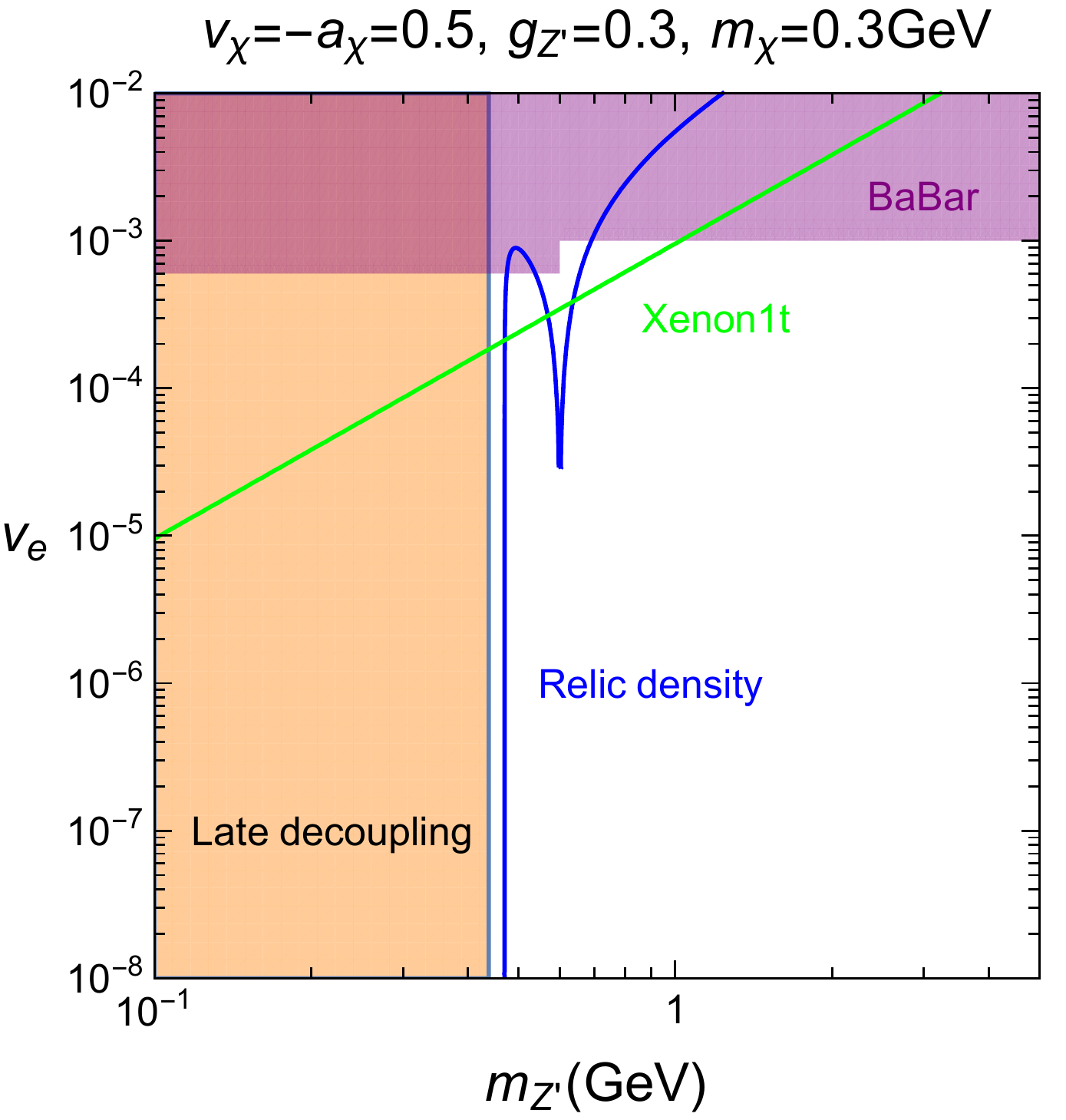}\,\,\,\,
\includegraphics[width=.40\textwidth]{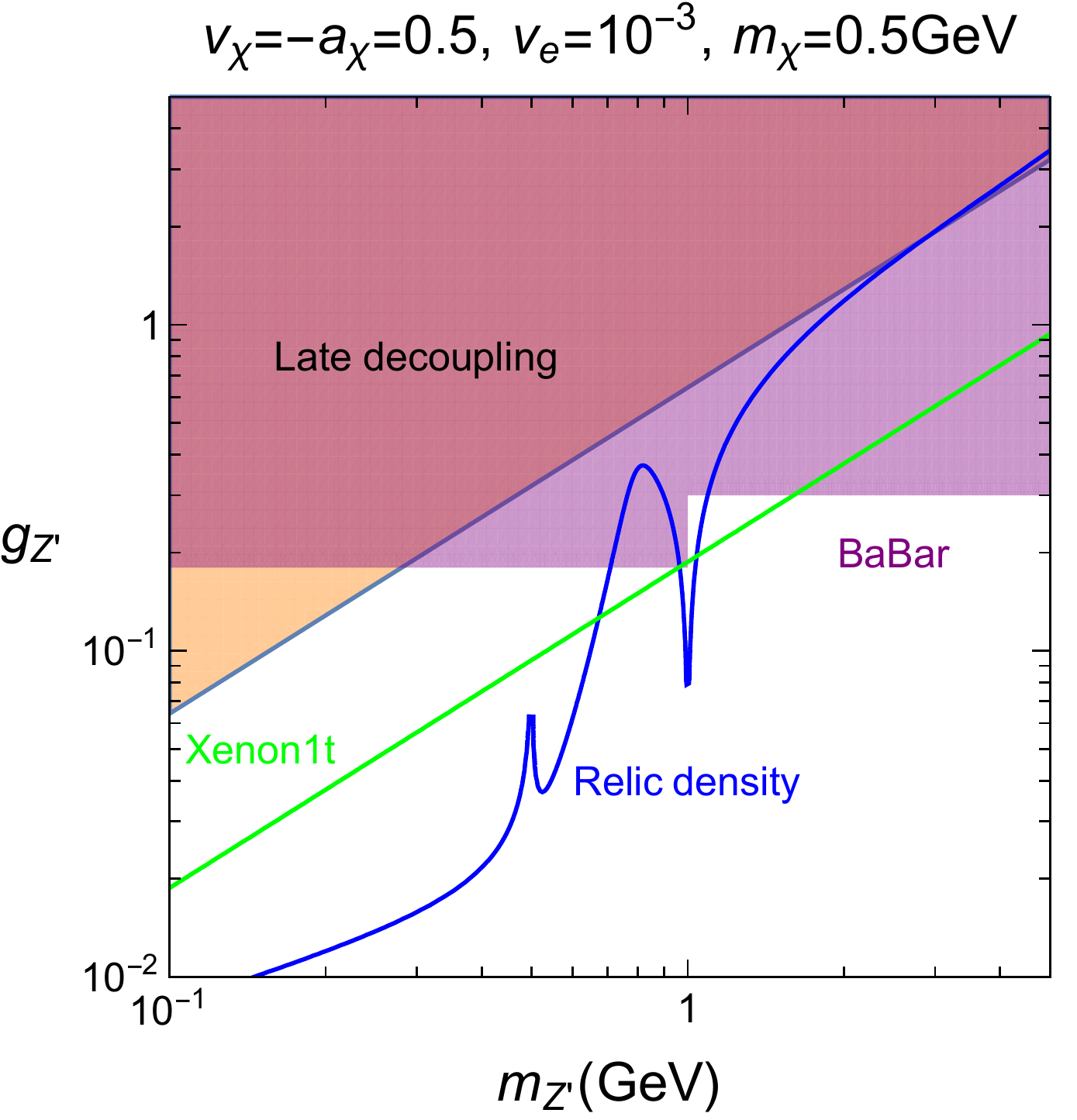}
  \caption{Parameter space for relic density and XENON1T electron excess.  The favored region by XENON1T electron excess is along the green line, the relic density is satisfied along the blue line. The purple region on the left plot is excluded by the BaBar limits on the visible and invisible decays of $Z'$. We have taken $v_\chi=-a_\chi=0.5$, $a_e=0$, and $m_\chi=0.3, 0.5\,{\rm GeV}$ on left and right plots, respectively, and  $g_{Z'}=0.3$ on left and $v_e=10^{-3}$ on right.  In the orange region, the relic density for the heavier state is Boltzmann-suppressed for $T_{\rm kd}=1\,{\rm MeV}$.}
  \label{relic}
\end{figure}

As a consequence, the dark matter number density is given by $n_{\rm DM}=n_{\chi_1}+n_{\chi_2}$ with $n_{\chi_1}\simeq n_{\chi_2}$ for the early decoupling of the $2\to 2 $ annihilations, so the corresponding relic abundance is determined as
\bea 
\Omega_{\rm DM} h^2=0.12\left( \frac{10.75}{g_*(T_f)}\right)^{1/2} \Big(\frac{x_f}{20} \Big) \left(\frac{4.3\times 10^{-9}\,{\rm GeV}^{-2}}{x_f \int^\infty_{x_f} x^{-2} \langle\sigma v\rangle} \right)
\eea
where $\langle\sigma v\rangle$ is given by eq.~(\ref{totalann}) and $x_f=m_{\chi_1}/T_f$ at freeze-out temperature.
Therefore, parametrizing the effective annihilation cross section by $\langle\sigma v\rangle=\frac{\alpha^2_{\rm eff}}{m^2_{\chi_1}}$, we can achieve a correct relic density, provided that
\bea
m_{\chi_1}\simeq 150\,{\rm MeV}\, \bigg(\frac{\alpha_{\rm eff}}{10^{-5}}\bigg).
\eea
As a result, we can get the correct relic density by taking a small effective coupling $\alpha_{\rm eff}$ from small SM couplings, $v_e, v_\nu$, or due to the Boltzmann-suppression, $ \alpha_{\rm eff}\sim g^2_{Z'}\, e^{-x_f(m_{Z'}-m_{\chi_1})}$, for $g_{Z'}=0.6$ and $\frac{m_{Z'}}{m_{\chi_1}}-1\simeq 0.4$, being consistent with the constraint from the Xenon excess with $\Omega_{\chi_1}\simeq \frac{1}{2} \Omega_{\rm DM}$ in eq.~(\ref{xenon}).

In Fig.~\ref{relic}, we show the parameter space for relic density in comparison to Xenon electron excess, for $m_{Z'}$ vs $v_e$ on left and $m_{Z'}$ vs $g_{Z'}$ on right. We have fixed $v_\chi=-a_\chi=0.5$, $a_e=0$, and $m_\chi=0.3, 0.5\,{\rm GeV}$ on left and right plots, respectively, and $g_{Z'}=0.3$ on left and $v_e=10^{-3}$ on right. We have imposed the BaBar bounds from visible and invisible decays of $Z'$ on the effective electron coupling $v_e$. The correct relic density is saturated along the blue lines, showing that there is a parameter space compatible with the Xenon electron excess in green lines and the BaBar bounds as well as the late decoupling in orange region.

In the above discussion, we focused on  the annihilation channels in the minimal scenario for the Xenon electron excess. 
However, if the aforementioned annihilation channels with $Z'$ interactions are not sufficient for a correct relic density, due to small $Z'$ couplings to the SM, we can also consider the dark matter self-interactions for dark matter annihilation, in particular, as in  the case of SIMP dark matter \cite{SIMP,SIMP2} where sub-GeV light dark matter is a natural outcome of the $3\rightarrow 2$ annihilations with strong self-interactions.
In this case, the relic density condition on the visible coupling becomes weaker, so there is more parameter opening up for the Xenon electron excess.

\section{Microscopic models}

In this section, we propose a microscopic model for exothermic dark matter by taking two left-handed  dark fermions, $\psi_1$ and $\psi_2$,  with opposite charges, $+1$ and $-1$, under the dark $U(1)'$ symmetry. 
We also introduce a dark Higgs $\phi$ with charge $-2$ under the $U(1)'$.
We assume that all the SM particles are neutral under the $U(1)'$, but dark matter can communicate with the SM through 1) the gauge kinetic mixing, $\sin\xi$, or 2) the mixing between electron and an extra vector-like lepton.
We first discuss the dark matter interactions and proceed to derive the effective interactions for the electron in each case of portal models. 

The Lagrangian for the dark sector is given, as follows,
\bea
{\cal L} &=&-\frac{1}{4} F'_{\mu\nu} F^{\prime \mu\nu} +|D_\mu\phi|^2 - V(\phi,H) \nonumber \\
&&+ i{\bar\psi}_{1L}\gamma^\mu  D_\mu \psi_{1L} + i{\bar\psi}_{2L}\gamma^\mu  D_\mu \psi_{2L}  \nonumber \\
&&-m_\psi \psi_1 \psi_2 - y_1\phi \, \psi_1 \psi_1 -y_2 \phi^* \psi_2 \psi_2 +{\rm h.c.}
\eea
where $F'_{\mu\nu}=\partial_\mu Z'_\nu-\partial_\nu Z'_\mu$, $B_{\mu\nu}$ is the field strength tensor for the SM hypercharge, the covariant derivatives are $D_\mu\phi=(\partial_\mu +2i g_{Z'} Z'_\mu)\phi$, $D_\mu \psi_{1L} =(\partial_\mu -ig_{Z'} Z'_\mu) \psi_{1L}$, $D_\mu \psi_{2L} =(\partial_\mu +ig_{Z'} Z'_\mu) \psi_{2L}$,  $m_\psi$ is the Dirac mass for dark fermions, and $y_{1,2}$ are the Yukawa couplings for dark fermions.
Here, $V(\phi,H)$ is the scalar potential for the singlet scalar $\phi$ and the SM Higgs.

After the dark Higgs gets a VEV as $\langle\phi\rangle=v_\phi$,  the $Z'$ gauge boson receives mass $m_{Z'}=2\sqrt{2} g_{Z'} v_\phi$, and  there appears a mass  mixing between $\psi_1$ and $\psi_2$. Then, diagonalizing the mass matrix for the dark fermions, we get the mass eigenvalues and the mixing matrix, as follows,
\bea
m^2_{\chi_{1,2}} =m^2_\psi +2 (y^2_1+y^2_2) v^2_\phi \pm 2\sqrt{(y^2_1-y^2_2)^2 v^4_\phi+ (y_1+y_2)^2 v^2_\phi m^2_\psi},
\eea 
and
\bea
\left(\begin{array}{c}  \chi_1  \\ \chi_2 \end{array} \right) =\left(\begin{array}{cc}  \cos\theta & -\sin\theta  \\ \sin\theta & \cos\theta \end{array} \right) \left(\begin{array}{c}  \psi_2  \\ \psi_1 \end{array} \right)
\eea
with
\bea
\sin2\theta= -\frac{4(y_1+y_2) v_\phi m_\psi}{m^2_{\chi_2}-m^2_{\chi_1}}.
\eea
For simplicity, we take $y_1=y_2$, then the mass eigenvalues become $m_{\chi_{1,2}} =m_\psi\pm 2y_1 v_\phi$ and the mixing angle is  given by $\theta=\frac{\pi}{4}$. Then, the mass difference can be small as far as $2|y_1| v_\phi\ll m_\psi$.
There was a recent discussion on the microscopic model for exothermic dark matter with a complex scalar field where the breaking of dark $U(1)'$ makes a small mass splitting between the real scalar fields \cite{harigaya}.

For instance, for $m_\psi\sim 1\,{\rm GeV}$, we need $y_1\sim 1.5\times 10^{-6}$.
As a result, including the $Z'$-portal couplings to the SM fermions for a small gauge kinetic mixing,  we summarize the $Z'$ gauge interactions as follows,
\bea
{\cal L}_{\rm DM} = -g_{Z'} Z'_\mu \Big( {\bar\chi}_{1} \gamma^\mu P_L \chi_{2} + {\bar\chi}_{2} \gamma^\mu  P_L\chi_{1} \Big).
\eea
Then, we obtain the effective dark matter couplings  in the Lagrangian (\ref{eft}) as
\bea
v_\chi=-a_\chi =  -\frac{1}{2}.
\eea
As a result, we can realize the transition interactions between two states of dark matter via the $Z'$ mediator, that are necessary for explaining the Xenon excess.

\begin{figure}[tbp]
  \centering
\includegraphics[width=.40\textwidth]{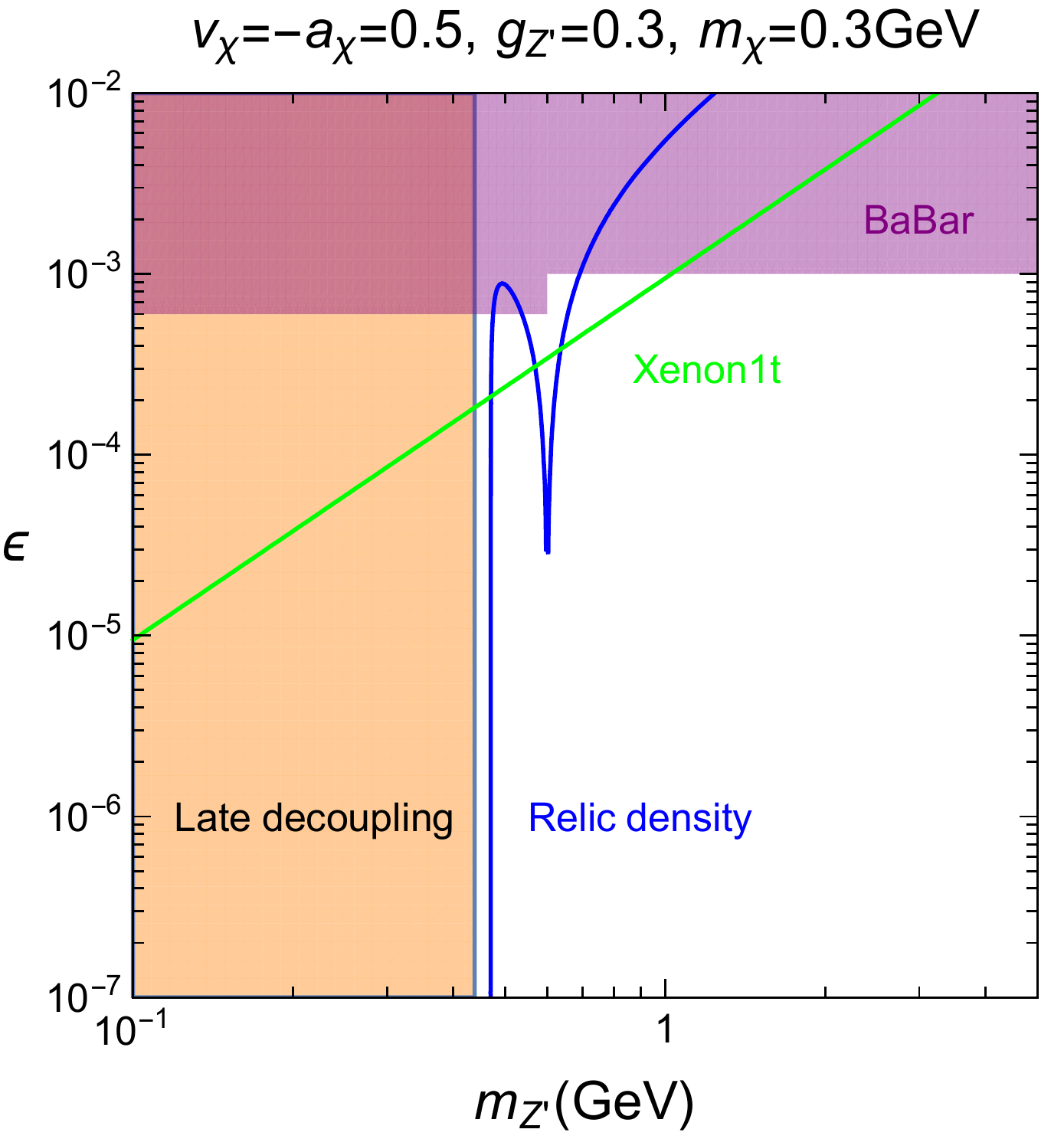} \,\,\,\,
\includegraphics[width=.40\textwidth]{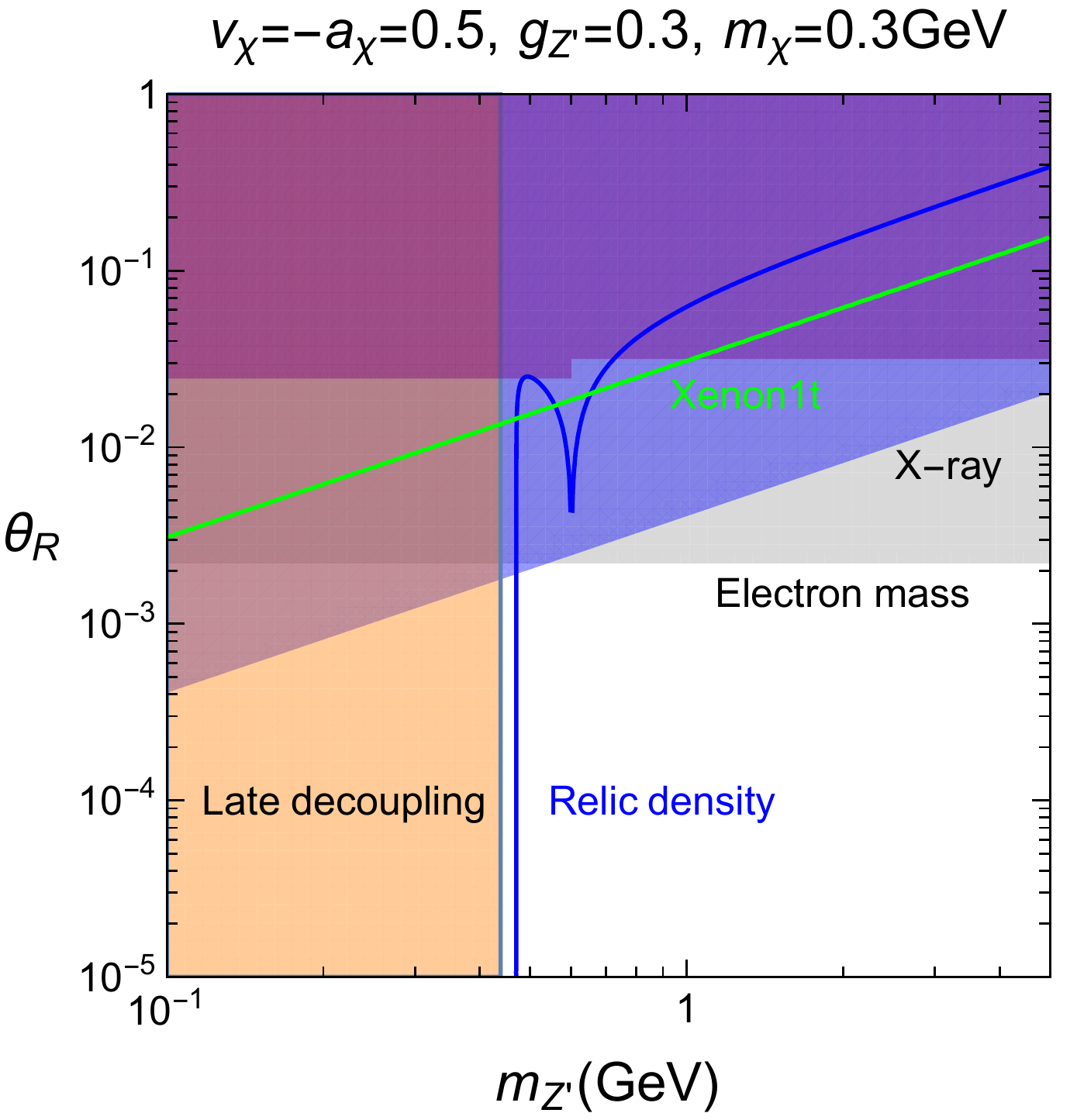}
  \caption{ Parameter space for relic density and XENON1T electron excess in microscopic models. (Left) $Z'$ portal: We have taken  $v_\chi=-a_\chi=0.5$,   $g_{Z'}=0.3$  and $m_\chi=0.3\,{\rm GeV}$. The color notations are the same as those in Fig.~\ref{relic}. (Right) Vector-like lepton portal: We have taken $v_\chi=-a_\chi=0.5$,   $g_{Z'}=0.3$  and $m_\chi=0.3\,{\rm GeV}$. The red region is excluded by the X-ray searches and the gray region is disfavored by the electron mass.  In the orange region, the relic density for the heavier state is Boltzmann-suppressed  for $T_{\rm kd}=1\,{\rm MeV}$.}
  \label{models}
\end{figure}

\subsection{$Z'$-portal}

In the presence of a gauge kinetic mixing,  
\bea
{\cal L}_{\rm kin-mix} =-\frac{1}{2} \sin\xi B_{\mu\nu} F^{\prime\mu\nu}, 
\eea
the mixing between $Z'$  and $Z$ gauge bosons gives rise to the $Z'$ gauge interactions to the SM as
\bea
{\cal L}_{\rm eff, I} = -  e\varepsilon\, Z'_\mu\bigg(  {\bar e} \gamma^\mu  e +  \frac{m^2_{Z'}}{2 c^2_W m^2_Z}\, {\bar\nu} \gamma^\mu P_L \nu \bigg)+\cdots
\eea
where $\varepsilon\equiv \xi \cos\theta_W$ and the ellipse denotes the electromagnetic and neutral current interactions for the rest of the SM fermions. 
Therefore, there are not only electron couplings but also neutrino couplings, although the latter being further suppressed by $m^2_{Z'}/m^2_Z$ \cite{z3dm}.
Consequently, we can identify the effective couplings in the Lagrangian (\ref{eft}), as follows,
\bea
 v_e=-\frac{e\varepsilon}{g_{Z'}}, \quad a_e=0,\quad\quad  v_\nu =-a_\nu= -\frac{e\varepsilon m^2_{Z'}}{4 c^2_W g_{Z'} m^2_Z }.
\eea
Then, the lifetime of the dark fermion $\chi_1$ is given by
\bea
\tau_{\chi_1} =\frac{1}{\Gamma(\chi_1\rightarrow \chi_2 \nu {\bar\nu})}= \bigg(\frac{10^{-4}e}{\varepsilon g_{Z'} }\bigg)^2\bigg(\frac{2.5\,{\rm keV}}{\Delta m} \bigg)^5 \, 8.9\times 10^{24}\,{\rm sec}.
\eea
Therefore, the dark fermion $\chi_1$ is much long-lived than the age of the Universe, so it can be responsible for the Xenon electron excess through the inelastic scattering, as discussed in the previous section. 

On the left of Fig.~(\ref{models}), we show the parameter space in $\varepsilon$ vs $m_{Z'}$, that is consistent with the Xenon electron excess, the correct relic density as well as the BaBar bounds. We have taken   $v_\chi=-a_\chi=0.5$,   $g_{Z'}=0.3$  and $m_\chi\equiv m_{\chi_1}=0.3\,{\rm GeV}$. Since the effective axial coupling $a_e$ vanishes, there is no bound from the X-ray searches.  In the orange region, the relic density for the heavier state is Boltzmann-suppressed due to the $2\to 2$ self-annihilations. As a result, there is a consistent parameter space for the Xenon electron excess, which can be probed in the future experiments.

\subsection{Vector-like lepton portal}

We introduce an extra vector-like charged lepton $E$ which has charge $-2$ under the $U(1)'$ but is singlet under the $SU(2)_L$.
Then, the mixing between the SM right-handed electron and the vector-like lepton is given by
\bea
{\cal L}_{\rm VL} = -M_E {\bar E} E - (y_E \phi {\bar E} e_R +{\rm h.c.})
\eea
As a consequence, the mass matrix for the electron and the vector-like lepton takes
\bea
M_e= \left(\begin{array}{cc} m_e & 0  \\ y_E v_\phi  & M_E  \end{array} \right).
\eea
Then, the mass eigenvalues are given by
\bea
m^2_{f_{1,2}} =\frac{1}{2} \bigg(m^2_e + M^2_E + y^2_E v^2_\phi \mp \sqrt{(m^2_e+y^2_E v^2_\phi-M^2_E)^2+4 y^2_E v^2_\phi M^2_E} \bigg).
\eea
On the other hand, the mixing angles for the right-handed electrons and the left-handed electrons are given \cite{VL}, respectively, by
\bea
\sin(2\theta_R) &=&-\frac{2y_E v_\phi M_E}{m^2_{f_1}-m^2_{f_2}}, \\
\sin(2\theta_L) &=& \frac{m^2_e}{m_{f_1} m_{f_2}}\,\sin(2\theta_R).
\eea
Therefore, for $m_e, y_E v_\phi\ll M_E$ and $(y_E v_\phi/M_E)^2\lesssim (m_e/M_E)$, the mass eigenvalues are approximated to $m_{f_1}\sim m_e$ and $m_{f_2}\sim M_E$, and the mixing angles become $\theta_R\sim \frac{2y_E v_\phi}{M_E}$ and $\theta_L\sim \frac{m_e}{M_E}\,\theta_R$.
Given the experimental bound on the vector-like charged lepton from LEP and LHC,  $M_E\gtrsim 100\,{\rm GeV}$, we have $\frac{m_e}{M_E}\lesssim 5\times 10^{-6}$, so we can ignore the mixing for the left-handed electrons. 
But, for $m_{f_1}\sim m_e$,  the mixing angle for the right-handed electrons is bounded by $\theta_R\lesssim \sqrt{\frac{m_e}{M_E}}$, thus $\theta_R$ can be as large as $2.2\times 10^{-3}$. 
On the other hand, we note that the suppressed mixing angle $\theta_L$ for the left-handed electron is consistent with electroweak precision data.

Consequently, we get  the following effective interactions for $Z'$,
\bea
{\cal L}_{\rm eff,II} &=&-2 g_{Z'} Z'_\mu \Big({\bar E} \gamma^\mu  E+ \theta^2_R\,  {\bar e} \gamma^\mu P_R e -\theta_R {\bar E}\gamma^\mu P_R e -\theta_R\, {\bar e}\gamma^\mu P_R E \nonumber \\
&&\quad+\theta^2_L {\bar e} \gamma^\mu P_L e -\theta_L  {\bar E}\gamma^\mu P_L e -\theta_L\, {\bar e}\gamma^\mu P_L E   \Big) \nonumber \\
&&-\frac{g}{2c_W}\, Z_\mu \Big( {\bar e}_L \gamma^\mu P_L e + \theta_L {\bar E}\gamma^\mu P_L e+\theta_L\,  {\bar e}\gamma^\mu P_L E+\theta^2_L{\bar E} \gamma^\mu P_L E    \Big) \nonumber \\
&&-\frac{g}{\sqrt{2}}\, \theta_L {\bar E} \gamma^\mu P_L \nu W^-_\mu +{\rm h.c.}
\eea
Here, we used the same notations for mass eigenstates as for interaction eigenstates, electron-like and vector-like. 
Therefore, we can identify the effective couplings in the Lagrangian (\ref{eft}), as follows,
\bea
v_e=a_e =  - \theta^2_R, \quad\quad v_\nu=a_\nu=0.
\eea
As a result, we obtain the necessary electron coupling to $Z'$ for explaining the Xenon excess through the small mixing between the SM right-handed electron and the vector-like lepton.
In the model with vector-like lepton portal, however, there is no direct coupling between $Z'$ and neutrinos. Even the $Z'-e-E$ vertex with $E$ decaying into the SM particles does not make the dark fermion $\chi_1$ to  decay, provided that $\Delta m< 2m_e$ is chosen.

On the right of Fig.~(\ref{models}), we show the parameter space in $\theta_R$ vs $m_{Z'}$, that is consistent with the Xenon electron excess, the correct relic density as well as the BaBar bounds. We have taken   $v_\chi=-a_\chi=0.5$,   $g_{Z'}=0.3$  and $m_\chi=m_{\chi_1}=0.3\,{\rm GeV}$. Since the effective axial coupling $a_e$ is nonzero, the bound from the X-ray searches in gray excludes the relic density condition in blue line, and  the consistent parameter space is pushed towards a larger $g_{Z'}$ or the resonance region near $m_{Z'}\sim 2m_\chi$. We note that the gray region is disfavored by the electron mass. We also note that in the orange region, the relic density for the heavier state is Boltzmann-suppressed due to the $2\to 2$ self-annihilations.

\section{Conclusions}

We proposed exothermic dark matter to explain the recent electron excess reported by XENON1T experiment.
 Even for a small dark matter velocity, as known from the standard Maxwellian  velocity distribution of dark matter, we achieved the appropriate recoil electron recoil energy at about $2.5\,{\rm keV}$ by considering the down-scattering of the heavier dark matter state  off the electron into the lighter state. Thus, we showed that about 50 recoil events over $E_R=2-3\,{\rm keV}$, which are most significant, can be explained in this scenario up to the detector resolution.
 
Including the effects of the phase-space enhancement  and the atomic excitation factor, we derived the required scattering cross section for the Xenon excess to be about $\sigma_e\sim 10^{-44}\,{\rm cm}^2$ for sub-GeV light dark matter. We took the effective theory approach for exothermic dark matter with a massive $Z'$ mediator and discussed the implications of the Xenon excess for dark matter interactions to $Z'$ and the dark matter relic density. We also provided microscopic models with $Z'$ portal and vector-like portal, realizing the required dark matter and electron couplings to $Z'$, while the heavier state of dark matter is long-lived enough to satisfy the bounds from the X-ray or the neutrino experiments.

\section*{Acknowledgments}

I would like to thank the CERN Theory group members for the brain-storming discussion on the Xenon excess during the virtual coffee meeting, Juri Smirnov, Jure Zupan and Bin Zhu for communications on the interpretation of the Xenon excess, and Yoo-Jin Kang and Soonbin Kim for checking the formulas for the electron recoil energy in the paper.
In particular, I appreciate the comment from Jure Zupan to check the two-photon decay mode of the heavier state of dark matter in more detail.
The work is supported in part by Basic Science Research Program
through the National Research Foundation of Korea (NRF) funded by the
Ministry of Education, Science and Technology (NRF-2019R1A2C2003738 and NRF-2018R1A4A1025334).




\begin{thebibliography}{999}



\bibitem{xenon}
E.~Aprile \textit{et al.} [XENON],
[arXiv:2006.09721 [hep-ex]].


\bibitem{fumi}
F.~Takahashi, M.~Yamada and W.~Yin,
[arXiv:2006.10035 [hep-ph]].


\bibitem{raidal}
K.~Kannike, M.~Raidal, H.~Veermäe, A.~Strumia and D.~Teresi,
[arXiv:2006.10735 [hep-ph]].


\bibitem{darkphoton}
G.~Alonso-Álvarez, F.~Ertas, J.~Jaeckel, F.~Kahlhoefer and L.~Thormaehlen,
[arXiv:2006.11243 [hep-ph]].



\bibitem{harigaya}
K.~Harigaya, Y.~Nakai and M.~Suzuki,
[arXiv:2006.11938 [hep-ph]].


\bibitem{rest}
B.~Fornal, P.~Sandick, J.~Shu, M.~Su and Y.~Zhao,
[arXiv:2006.11264 [hep-ph]];
C.~Boehm, D.~G.~Cerdeno, M.~Fairbairn, P.~A.~Machado and A.~C.~Vincent,
[arXiv:2006.11250 [hep-ph]];
L.~Su, W.~Wang, L.~Wu, J.~M.~Yang and B.~Zhu,
[arXiv:2006.11837 [hep-ph]].
M.~Du, J.~Liang, Z.~Liu, V.~Tran and Y.~Xue,
[arXiv:2006.11949 [hep-ph]];
L.~Di Luzio, M.~Fedele, M.~Giannotti, F.~Mescia and E.~Nardi,
[arXiv:2006.12487 [hep-ph]];
N.~F.~Bell, J.~B.~Dent, B.~Dutta, S.~Ghosh, J.~Kumar and J.~L.~Newstead,
[arXiv:2006.12461 [hep-ph]];
Y.~Chen, J.~Shu, X.~Xue, G.~Yuan and Q.~Yuan,
[arXiv:2006.12447 [hep-ph]].
D.~Aristizabal Sierra, V.~De Romeri, L.~Flores and D.~Papoulias,
[arXiv:2006.12457 [hep-ph]];
J.~Buch, M.~A.~Buen-Abad, J.~Fan and J.~S.~C.~Leung,
[arXiv:2006.12488 [hep-ph]];
G.~Choi, M.~Suzuki and T.~T.~Yanagida,
[arXiv:2006.12348 [hep-ph]];
G.~Paz, A.~A.~Petrov, M.~Tammaro and J.~Zupan,
[arXiv:2006.12462 [hep-ph]];
A.~Bally, S.~Jana and A.~Trautner,
[arXiv:2006.11919 [hep-ph]].



\bibitem{exothermic}
B.~Batell, M.~Pospelov and A.~Ritz,
Phys. Rev. D \textbf{79} (2009), 115019
doi:10.1103/PhysRevD.79.115019
[arXiv:0903.3396 [hep-ph]];
P.~W.~Graham, R.~Harnik, S.~Rajendran and P.~Saraswat,
Phys. Rev. D \textbf{82} (2010), 063512
doi:10.1103/PhysRevD.82.063512
[arXiv:1004.0937 [hep-ph]];
R.~Essig, J.~Kaplan, P.~Schuster and N.~Toro,
[arXiv:1004.0691 [hep-ph]].


\bibitem{3to2}
J.~Smirnov and J.~F.~Beacom,
[arXiv:2002.04038 [hep-ph]].


\bibitem{general}
R.~Essig, M.~Fernandez-Serra, J.~Mardon, A.~Soto, T.~Volansky and T.~T.~Yu,
JHEP \textbf{05} (2016), 046
doi:10.1007/JHEP05(2016)046
[arXiv:1509.01598 [hep-ph]].


\bibitem{excitation}
B.~Roberts, V.~Dzuba, V.~Flambaum, M.~Pospelov and Y.~Stadnik,
Phys. Rev. D \textbf{93} (2016) no.11, 115037
doi:10.1103/PhysRevD.93.115037
[arXiv:1604.04559 [hep-ph]].


\bibitem{ionization}
R.~Essig, J.~Mardon and T.~Volansky,
Phys. Rev. D \textbf{85} (2012), 076007
doi:10.1103/PhysRevD.85.076007
[arXiv:1108.5383 [hep-ph]];
R.~Essig, A.~Manalaysay, J.~Mardon, P.~Sorensen and T.~Volansky,
Phys. Rev. Lett. \textbf{109} (2012), 021301
doi:10.1103/PhysRevLett.109.021301
[arXiv:1206.2644 [astro-ph.CO]];
R.~Essig, T.~Volansky and T.~T.~Yu,
Phys. Rev. D \textbf{96} (2017) no.4, 043017
doi:10.1103/PhysRevD.96.043017
[arXiv:1703.00910 [hep-ph]].


\bibitem{DD}
G.~Jungman, M.~Kamionkowski and K.~Griest,
Phys. Rept. \textbf{267} (1996), 195-373
doi:10.1016/0370-1573(95)00058-5
[arXiv:hep-ph/9506380 [hep-ph]];
A.~Carrillo-Monteverde, Y.~J.~Kang, H.~M.~Lee, M.~Park and V.~Sanz,
JHEP \textbf{06} (2018), 037
doi:10.1007/JHEP06(2018)037
[arXiv:1803.02144 [hep-ph]].


\bibitem{fdm}
K.~Griest and D.~Seckel,
Phys. Rev. D \textbf{43} (1991), 3191-3203
doi:10.1103/PhysRevD.43.3191;
R.~T.~D'Agnolo and J.~T.~Ruderman,
Phys.\ Rev.\ Lett.\  {\bf 115} (2015) no.6,  061301
doi:10.1103/PhysRevLett.115.061301
[arXiv:1505.07107 [hep-ph]];
S.~M.~Choi, Y.~J.~Kang and H.~M.~Lee,
JHEP \textbf{12} (2016), 099
doi:10.1007/JHEP12(2016)099
[arXiv:1610.04748 [hep-ph]];
Y.~J.~Kang and H.~M.~Lee,
[arXiv:2001.04868 [hep-ph]];
S.~M.~Choi, J.~Kim, H.~M.~Lee and B.~Zhu,
[arXiv:2003.11823 [hep-ph]].


\bibitem{cbpark}
H.~M.~Lee, C.~B.~Park and M.~Park,
Phys. Lett. B \textbf{744} (2015), 218-224
doi:10.1016/j.physletb.2015.03.046
[arXiv:1501.05479 [hep-ph]].


\bibitem{tait}
C.~Jackson, G.~Servant, G.~Shaughnessy, T.~Tait, M.P. and M.~Taoso,
JCAP \textbf{07} (2013), 021
doi:10.1088/1475-7516/2013/07/021
[arXiv:1302.1802 [hep-ph]];
C.~Jackson, G.~Servant, G.~Shaughnessy, T.~Tait, M.P. and M.~Taoso,
JCAP \textbf{07} (2013), 006
doi:10.1088/1475-7516/2013/07/006
[arXiv:1303.4717 [hep-ph]].


\bibitem{xray}
M.~Drewes et al,
JCAP \textbf{01} (2017), 025
doi:10.1088/1475-7516/2017/01/025
[arXiv:1602.04816 [hep-ph]].


\bibitem{SK}
S.~Desai \textit{et al.} [Super-Kamiokande],
Phys. Rev. D \textbf{70} (2004), 083523
doi:10.1103/PhysRevD.70.083523
[arXiv:hep-ex/0404025 [hep-ex]].

\bibitem{ibarra}
L.~Covi, M.~Grefe, A.~Ibarra and D.~Tran,
JCAP \textbf{04} (2010), 017
doi:10.1088/1475-7516/2010/04/017
[arXiv:0912.3521 [hep-ph]];
H.~Zhang \textit{et al.} [Super-Kamiokande],
Astropart. Phys. \textbf{60} (2015), 41-46
doi:10.1016/j.astropartphys.2014.05.004
[arXiv:1311.3738 [hep-ex]];
E.~Richard \textit{et al.} [Super-Kamiokande],
Phys. Rev. D \textbf{94} (2016) no.5, 052001
doi:10.1103/PhysRevD.94.052001
[arXiv:1510.08127 [hep-ex]];
P.~Bandyopadhyay, E.~J.~Chun and R.~Mandal,
JCAP \textbf{08} (2020), 019
doi:10.1088/1475-7516/2020/08/019
[arXiv:2005.13933 [hep-ph]].


\bibitem{babar}
J.~Lees \textit{et al.} [BaBar],
Phys. Rev. Lett. \textbf{113} (2014) no.20, 201801
doi:10.1103/PhysRevLett.113.201801
[arXiv:1406.2980 [hep-ex]].


\bibitem{babarinv}
J.~Lees \textit{et al.} [BaBar],
Phys. Rev. Lett. \textbf{119} (2017) no.13, 131804
doi:10.1103/PhysRevLett.119.131804
[arXiv:1702.03327 [hep-ex]].


\bibitem{beamdump}
  J.~D.~Bjorken, S.~Ecklund, W.~R.~Nelson, A.~Abashian, C.~Church, B.~Lu, L.~W.~Mo and T.~A.~Nunamaker {\it et al.},
  Phys.\ Rev.\ D {\bf 38} (1988) 3375;
  B.~Batell, R.~Essig and Z.~Surujon,
  Phys.\ Rev.\ Lett.\  {\bf 113} (2014) 17,  171802
  [arXiv:1406.2698 [hep-ph]];
D.~Banerjee \textit{et al.} [NA64],
Phys. Rev. Lett. \textbf{118} (2017) no.1, 011802
doi:10.1103/PhysRevLett.118.011802
[arXiv:1610.02988 [hep-ex]];
D.~Banerjee, V.~E.~Burtsev, A.~G.~Chumakov, D.~Cooke, P.~Crivelli, E.~Depero, A.~V.~Dermenev, S.~V.~Donskov, R.~R.~Dusaev and T.~Enik, \textit{et al.}
Phys. Rev. Lett. \textbf{123} (2019) no.12, 121801
doi:10.1103/PhysRevLett.123.121801
[arXiv:1906.00176 [hep-ex]].


\bibitem{SIMP}
  Y.~Hochberg, E.~Kuflik, T.~Volansky and J.~G.~Wacker,
  Phys.\ Rev.\ Lett.\  {\bf 113} (2014) 171301
  doi:10.1103/PhysRevLett.113.171301
  [arXiv:1402.5143 [hep-ph]];
  Y.~Hochberg, E.~Kuflik, H.~Murayama, T.~Volansky and J.~G.~Wacker,
  Phys.\ Rev.\ Lett.\  {\bf 115} (2015) no.2,  021301
  doi:10.1103/PhysRevLett.115.021301
  [arXiv:1411.3727 [hep-ph]];
  H.~M.~Lee and M.~S.~Seo,
  Phys.\ Lett.\ B {\bf 748} (2015) 316
  doi:10.1016/j.physletb.2015.07.013
  [arXiv:1504.00745 [hep-ph]].
  
  
\bibitem{SIMP2}
S.~M.~Choi, H.~M.~Lee and M.~S.~Seo,
JHEP \textbf{04} (2017), 154
doi:10.1007/JHEP04(2017)154
[arXiv:1702.07860 [hep-ph]];
S.~M.~Choi, Y.~Hochberg, E.~Kuflik, H.~M.~Lee, Y.~Mambrini, H.~Murayama and M.~Pierre,
JHEP \textbf{10} (2017), 162
doi:10.1007/JHEP10(2017)162
[arXiv:1707.01434 [hep-ph]];
S.~M.~Choi, H.~M.~Lee, P.~Ko and A.~Natale,
Phys. Rev. D \textbf{98} (2018) no.1, 015034
doi:10.1103/PhysRevD.98.015034
[arXiv:1801.07726 [hep-ph]];
S.~M.~Choi, H.~M.~Lee, Y.~Mambrini and M.~Pierre,
JHEP \textbf{07} (2019), 049
doi:10.1007/JHEP07(2019)049
[arXiv:1904.04109 [hep-ph]].


\bibitem{z3dm}
S.~M.~Choi and H.~M.~Lee,
JHEP \textbf{09} (2015), 063
doi:10.1007/JHEP09(2015)063
[arXiv:1505.00960 [hep-ph]].


\bibitem{VL}
H.~M.~Lee, M.~Park and W.~I.~Park,
JHEP \textbf{12} (2012), 037
doi:10.1007/JHEP12(2012)037
[arXiv:1209.1955 [hep-ph]].


\end{thebibliography}
\end{document}